\documentclass{article}
\usepackage[preprint]{neurips_2025}

\usepackage{natbib} 
\usepackage[colorlinks=true, citecolor=blue, linkcolor=blue, urlcolor=blue, anchorcolor=blue]{hyperref}
\usepackage{graphicx}
\usepackage[utf8]{inputenc} 
\usepackage[T1]{fontenc}    
\usepackage{breakurl}     
\usepackage{listings}
\usepackage{caption}
\usepackage{booktabs}       
\usepackage{listings}
\usepackage{placeins}
\usepackage{amsfonts}       
\usepackage{amsmath}
\usepackage{amsthm}
\usepackage{xurl}
\usepackage{amssymb}
\usepackage{nicefrac}       
\usepackage{microtype}      
\usepackage[font=small,labelfont=bf]{caption}
\usepackage{enumitem}
\usepackage{wrapfig}
\usepackage{listings}
\usepackage{caption}
\usepackage{multirow}
\usepackage[most]{tcolorbox}
\usepackage[normalem]{ulem}
\usepackage{xspace}
\usepackage{float}
\usepackage{tabularx}
\usepackage{booktabs}
\usepackage[normalem]{ulem}
\usepackage{soul}   
\usepackage{svg}
\usepackage{array}
\usepackage{longtable}
\usepackage{adjustbox}
\usepackage{makecell}
\usepackage{pifont}

\usepackage{xcolor}

\useunder{\uline}{\ul}{}

\newcommand{\asr}{audio-to-text repetition}
\newcommand{\interleave}{cross-modal continuation}

\newcommand{\spnext}{\texttt{<next>}}
\newcommand{\sprepeat}{\texttt{<repeat>}}

\definecolor{mygreybg}{gray}{0.95}

\sethlcolor{mygreybg}

\title{Voxtral}

\begin{document}

\maketitle

\vspace{-0.1in}
\begin{center}
\vspace{-45pt}
\centering
\includegraphics[width=0.8\linewidth,keepaspectratio]{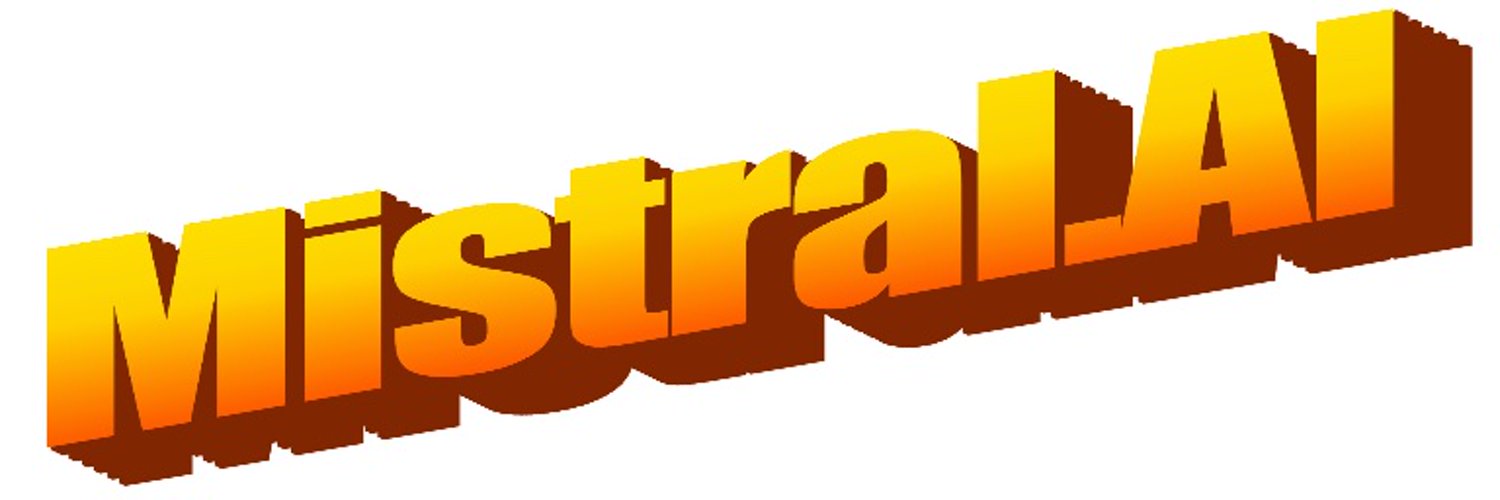}
\end{center}

\begin{abstract}

We present Voxtral Mini and Voxtral Small, two multimodal audio chat models. Voxtral is trained to comprehend both spoken audio and text documents, achieving state-of-the-art performance across a diverse range of audio benchmarks, while preserving strong text capabilities. Voxtral Small outperforms a number of closed-source models, while being small enough to run locally. A 32K context window enables the model to handle audio files up to 40 minutes in duration and long multi-turn conversations. We also contribute three benchmarks for evaluating speech understanding models on knowledge and trivia. Both Voxtral models are released under Apache 2.0 license.

\end{abstract}
\begin{center}
\begin{tabular}{@{} l l @{}}
\small{\textbf{Webpage:}}  & \scriptsize{\url{https://mistral.ai/news/voxtral/}} \\
\multirow{2}{*}{\small{\textbf{Model weights:}}}  & \scriptsize{\url{https://huggingface.co/mistralai/Voxtral-Mini-3B-2507}} \\
      & \scriptsize{\url{https://huggingface.co/mistralai/Voxtral-Small-24B-2507}} \\
\small{\textbf{Evals:}}      & \scriptsize{\url{https://huggingface.co/collections/mistralai/speech-evals-6875e9b26c78be4a081050f4}}
\end{tabular}
\end{center}

\vspace{-0.3cm}

\section{Introduction}
This paper describes Voxtral Mini and Voxtral Small, a pair of multimodal language models trained to understand both speech and text, released with open-weights under an Apache 2.0 license. Voxtral is pretrained on a large-scale corpus of audio and text documents, and subsequently instruction tuned on real and synthetic data. It is capable of responding directly to audio (or text) and answering questions about audio files. With a 32K token context window, Voxtral is capable of processing audio files up to 40 minutes long.

Compared with similarly sized models in the same evaluation setting, we find that Voxtral delivers strong audio reasoning capabilities without sacrificing text-only performance. Its performance is state-of-the-art for speech transcription and translation, outperforming other open-weights and closed models. In speech question-answering (QA) and summarization, it performs comparably with closed models of a similar price class, such as GPT-4o mini \citep{hurst2024gpt} and Gemini 2.5 Flash \citep{comanici2025gemini}.

During evaluation of Voxtral and other models, we found that the existing ecosystem of speech evaluations lacked breadth and standardization; the majority of previous work focused on evaluation of transcription and translation quality, and less on other understanding tasks. In Section~\ref{sec:synthetic-evals}, we present evaluations that measure a wider range of speech comprehension and reasoning tasks.

Our primary contributions are:
\begin{itemize}
    \item Two open-weights audio models with state-of-the-art transcription and multilingual speech understanding for audio durations up to their 32K context window\vspace{-0.3em}
    \item Native function calling support with audio\vspace{-0.3em}
    \item Evaluation benchmarks that measure speech understanding and reasoning
\end{itemize}

The report is structured as follows: First, we outline our modeling choices. Next, we describe methods for pretraining, post-training, and response quality enhancement. Finally, we present benchmark results along with architectural and data ablations.

\section{Modeling}
\begin{figure}[h]
    \centering
    \includegraphics[width=\textwidth]{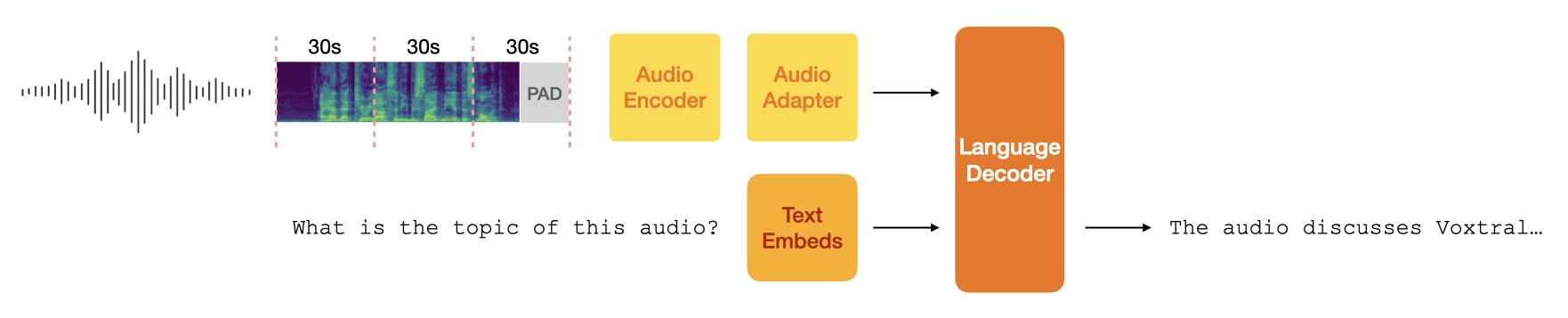}
    \caption{\textbf{Voxtral Architecture.} The audio encoder processes the speech input, attending to 30-second chunks of audio independently. The audio embeddings are concatenated at the output, and downsampled by a factor of 4x in the audio-language adapter. The multimodal LLM decoder auto-regressively predicts text tokens, conditional on the audio and text inputs.
    }
    \label{fig:architecture}
\end{figure}

Voxtral is based on the Transformer architecture \citep{vaswani2017attention}, consisting of three components: an audio encoder to process speech inputs, an adapter layer to downsample audio embeddings, and a language decoder to reason and generate text outputs. The overall architecture is depicted in Figure \ref{fig:architecture}.

\subsection{Audio Encoder}

The audio encoder is based on Whisper large-v3 \citep{radford2023whisper}. A raw audio waveform is first mapped to a log-Mel spectrogram \citep{davis1980comparison} with 128 Mel-bins and 160 hop-length.
Within the Whisper encoder, the spectrogram passes through a convolutional stem that downsamples its temporal resolution by a factor of two, after which it is fed into a stack of bidirectional self-attention layers. The resulting audio embeddings have a frame rate of 50 Hz.

Whisper has a fixed receptive field of 30 seconds. To accommodate audio sequences exceeding this duration, we compute the log-Mel spectrogram for the entire audio, but restrict the encoder to independently process each 30 second chunk. The absolute positional encodings are reset for each chunk, and chunks from the same audio are partitioned into a batch axis. Within the encoder's attention layers, this approach is functionally equivalent to chunk-wise attention \citep{zhang2023googleusmscalingautomatic}, which mitigates the computational overhead for longer audio inputs and enhances length generalization. The embeddings computed from each chunk are concatenated at the output stage, forming a unified representation of the complete audio sequence.

Due to its fixed receptive field, Whisper also pads short audios to 30 seconds. In Section \ref{sec:pad-analysis}, we investigate removing this padding requirement to allow continuous audio lengths. However, empirical results demonstrated a decline in performance, even when tuning the encoder to adapt. Consequently, we maintain the practice of padding all audio inputs to the next multiple of 30 seconds.

\subsection{Adapter Layer}

The high frame-rate of the audio encoder would result in long sequence-lengths through the language decoder. For example, a 30 minute audio at 50Hz has a sequence length of 90k tokens, leading to high memory and slow inference. To circumvent this, we append an additional MLP layer at the audio encoder outputs that is responsible for downsampling the audio embeddings. In Section \ref{sec:downsample-analysis}, we show a downsampling factor of 4x yields the best trade-off between sequence-length and performance. This results in an effective frame-rate of 12.5Hz, enabling Voxtral to gracefully handle audios up to 40 minutes with a context-length of 32k tokens.

\subsection{Language Decoder}

We release two variants of Voxtral: Mini and Small. Voxtral Mini is built on top of Ministral 3B \citep{mistral2024ministral}, an edge-focused model that delivers competitive performance with a small memory footprint. Voxtral Small leverages the Mistral Small 3.1 24B backbone \citep{mistral2024small3.1}, giving strong performance across a range of knowledge and reasoning tasks. Table \ref{tab:param-counts} decomposes the number of parameters in each checkpoint based on the sub-components.

\begin{table}[]
\centering
\caption{\textbf{Parameter Counts.} Number of parameters for Voxtral Mini and Small.} \label{tab:param-counts}
\begin{tabular}{@{}lrrrrr@{}}
\toprule
      & \multicolumn{1}{l}{Audio Encoder} & \multicolumn{1}{l}{Audio Adapter} & \multicolumn{1}{l}{Text Embeddings} & \multicolumn{1}{l}{Language Decoder} & \multicolumn{1}{l}{Total} \\ \midrule
Mini  & 640M                              & 25M                               & 400M                                & 3.6B                                 & 4.7B                      \\
Small & 640M                              & 52M                               & 670M                                & 22.9B                                & 24.3B                     \\ \bottomrule
\end{tabular}
\end{table}

\section{Methodology}
We train the model in three phases: pretraining, supervised finetuning, and preference alignment. Each phase is described separately below.
Finally, we describe our evaluation protocol for speech understanding tasks.

\subsection{Pretraining}
The pretraining stage of Voxtral is designed to introduce speech to the language decoder, complementary to the existing modality of text.  Given an audio dataset with text transcriptions, we first chunk the audio into short segments together with their corresponding transcription, forming parallel audio-text pairs: $(A_1,T_1), (A_2,T_2), \dots, (A_N,T_N)$. The segmentation boundaries are defined by upstream voice activity detection and diarization models. If transcripts are unavailable, we pseudo-label the audio with an ASR model.

Similar to prior works \citep{nguyen2025spirit,zeng2024scaling}, we define two patterns that combine audio and text into training samples for the model: \textit{\asr} and \textit{\interleave}.

The \asr~pattern is defined as an audio segment $A_n$ followed by the corresponding transcription $T_n$. A training sample consists of a single audio-text pair $(A_n, T_n)$. This formulation mimics speech recognition and is used to explicitly teach the model speech-to-text alignment. 

On the other hand, the \interleave~pattern is designed to implicitly align the speech and text modalities through modality-invariant context modeling. Specifically, for each audio segment $A_n$, the corresponding text is the proceeding text segment in the sequence $T_{n+1}$. In addition, a training sample is composed by interleaving audio and text for multiple consecutive segments: $(A_1, T_2, A_3, T_4, \dots, A_{N-1}, T_N)$. This structure resembles tasks like QA or conversation, where the model must maintain discourse continuity across modalities.

Since we use two different data patterns, the proceeding text segment for a given audio segment is ambiguous; both repeat and continuation are valid. To eliminate ambiguity, we introduce two special tokens to specify the expected output: \sprepeat~for repetition and \spnext~for continuation. These tokens are used for pattern indication during training and as part of the prompt during inference to control model behavior.

\begin{figure}
    \centering
    \includegraphics[width=\linewidth]{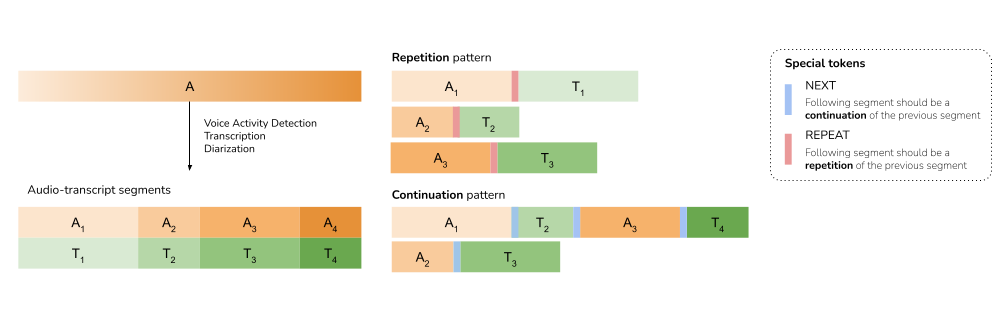}
    \caption{\textbf{Pretraining patterns}. A single audio-text example $(A,T)$ is first segmented into a set of audio-text pairs $\left\{(A_n, T_n)\right\}_{n=1}^{N}$, based on the timestamps and transcriptions returned by segmentation stage. For the \asr~pattern, a given audio $A_n$ is repeated in the text space $T_n$. For the \interleave~pattern, each audio $A_n$ is followed by its subsequent text $T_{n+1}$. The task is signaled to the model by the \sprepeat~ and \spnext~ special tokens respectively.}
    \label{fig:pretrain-pattern}
\end{figure}

The two pretraining patterns are shown with their special tokens in Figure~\ref{fig:pretrain-pattern}. Note that we treat each audio-transcription pair as a standalone sequence wrapped with \texttt{<bos>}/\texttt{<eos>} without previous context.

During pretraining, we balance the two patterns evenly. We demonstrate in Section \ref{sec:interleave-analysis} that this balanced approach is essential; the \asr~pattern drives transcription performance, while the \interleave~pattern prepares the model for speech understanding tasks that require deeper reasoning and context integration, such as audio-based QA or dialogue. To preserve text capabilities, we also include text pretraining data in the data mixture.

For the first pass over the data mixture, we freeze the audio encoder and language decoder, training only the adapter. We found this warm-up stage beneficial for speech understanding evaluations, whereas speech recognition results are similar with and without warm-up. We also perform one pretraining run on the Mini scale with just the \asr~pattern. We call this model Voxtral Mini Transcribe, and compare it to other ASR-only models in Section \ref{sec:asr-results}.

\subsection{Supervised Finetuning}
In post-training, our primary objective is to preserve or slightly enhance the transcription capabilities established during pretraining, while simultaneously extending the model’s proficiency over a range of speech understanding tasks. We also develop robust instruction-following behavior, irrespective of whether user inputs are in audio or text form.

Our speech understanding data falls into two categories: tasks where audio is provided as context and the assistant responds to text queries, and tasks where the assistant responds directly to audio inputs. Both categories rely significantly on synthetic data.

\paragraph{Audio Context, Text Query} To create synthetic data for tasks involving audio context paired with text queries, we utilize long-form audio data (segments up to approximately 40 minutes) with corresponding transcripts and language identification metadata. Transcripts are paired with tailored prompts and fed into an LLM (Mistral Large), which then generates question-answer pairs related to the audio content. The prompts explicitly instruct the LLM to frame both questions and answers as though they arise from auditory comprehension rather than text analysis, thereby encouraging natural responses from the downstream audio model. To achieve data diversity and richness, we vary question types, including straightforward factual inquiries, "needle-in-haystack" retrieval tasks, and reasoning-intensive problems. Moreover, to minimize repetitive question styles, the LLM generates multiple candidate question-answer pairs per audio segment, from which we sample a single pair for inclusion in the post-training dataset. While we typically ensure that the question-answer pairs match the language of the original audio and transcript, we occasionally instruct Mistral Large to produce pairs in different languages to enable QA for audios in languages the user does not speak.

Additionally, we allocate another portion of the long-form audio data for synthetic summarization and translation data. For translation tasks, we leverage language identification metadata to select a target language different from the original audio language. To mitigate overfitting to a narrow range of user message patterns, we sampled from a large, manually curated set of plausible user requests.

\paragraph{Audio-Only Input} For scenarios in which the user provides only audio input, we adapt existing text supervised finetuning data, including function calling datasets, by converting text user messages into synthetic audio using a text-to-speech (TTS) model. However, reliance solely on TTS-generated audio leads to poor generalization to genuine human speech, particularly accented voices, manifesting most commonly in erroneous transcription of conversational prompts rather than appropriate continuation. To address this limitation, we extract questions from long-form ASR data that can be adequately answered through general world knowledge, thus requiring no additional audio context. We then isolate audio excerpts containing these questions and generate corresponding text answers using Mistral Large. This process yields datasets consisting of genuine human speech questions paired with text answers. 

Speech recognition is a distinctive use case characterized by an unambiguous task, rendering the text prompt redundant. To address this, we introduce a dedicated "transcribe mode," signaled via a new special token. This mode explicitly instructs the model to perform transcription tasks, thereby eliminating the need for a text prompt.
\subsection{Preference Alignment}
\label{sec:response-quality}
Direct Preference Optimization (DPO) \citep{rafailov2024directpreferenceoptimizationlanguage} offers a lightweight alternative to full RLHF by learning directly from pairwise preferences. We also adopt its \textit{online} variant \citep{guo2024directlanguagemodelalignment}, where for each example, we sample two candidate responses from the current policy with temperature $T{=}0.5$. To rank responses, we take the entire conversation, replace the audio with its transcription, and leverage a text-based reward model. Although the reward model only has access to the audio transcription - rather than the raw audio itself - it is able to capture semantics, style, and factual coherence from this information, attributes that transfer to the generated text response. Our Online DPO implementation utilizes the sampling and reward infrastructure that powered the Magistral \citep{mistralai2025magistral} series.

We apply DPO and Online DPO to both Voxtral Mini and Small, for which we present results in Section \ref{sec:response-quality-results}. While both DPO and Online DPO helped improve the response quality, the online variant was more effective.

\subsection{Evaluation} \label{sec:synthetic-evals}
In addition to standard benchmarks for speech transcription, translation, and understanding - detailed in Sections \ref{sec:appendix-asr-full} and \ref{sec:appendix-au-full} - we create our own test sets. These sets build upon existing research and evaluate model attributes that are typically underrepresented, particularly long-context QA.

\paragraph{Speech-Synthesized Benchmarks} To evaluate spoken-language understanding, prior works take existing text benchmarks and synthesize the text prompt into speech \citep{nachmani2024spokenqa, chen2024voicebench}. We extend these test suites by creating speech-synthesized versions of three established text benchmarks: GSM8K \citep{gsm8k}, TriviaQA \citep{triviaqa}, and MMLU \citep{mmlu}.

The first step in creating these benchmarks involves filtering to only include prompts that are viable as speech-synthesized inputs, similar to \cite{llamaomni}. For every  example, we classify it into one of three categories with Mistral Large:

\begin{itemize}
  \item \textbf{Verbalizable}: plain wording or simple numerals. No re-write necessary.
  \item \textbf{Verbalizable with Rewrite}: math, code, or symbols, that can be deterministically rewritten into speech‑friendly text. For example, digits are converted to spelled-out form, acronyms expanded, markdown removed. The specific prompt used to achieve this is outlined in Appendix~\ref{sec:appendix-synthetic-eval-prompt}.
  \item \textbf{Non‑Verbalizable}: text that cannot be naturally spoken, such as tables, figures or lengthy math and code, is discarded.
\end{itemize}

Once the valid set of examples is established, we synthesize each one individually using a TTS engine. To ensure diversity in speakers, we randomly sample speaker embeddings from a diverse set, trimmed to six-second clips and filtered to only include single-speaker utterances. For each prompt, we sample a speaker embedding from this pool and generate the corresponding audio input using the TTS engine. Since the model output is in the text-space, scoring the generations requires no additional modifications.

We are releasing the synthesized evaluations under a permissive license and encourage their adoption as standard benchmarks for speech understanding.

\paragraph{Speech Understanding (SU) Benchmark} We develop an internal benchmark that measures the ability of models to answer questions about audios in a helpful manner. The audio files range up to 19 minutes in duration, assessing understanding on moderately long audio contexts. We use an LLM as a judge, which has access to a transcription of the audio, the question, a reference answer, and the proposed answer. The LLM judge returns two complementary metrics:

\begin{enumerate}
  \item \textbf{\textsc{llm\_judge\_score}}: a \emph{binary helpfulness indicator}.  
        The score is 1 if the answer is deemed correct and helpful to the user’s question, 0 otherwise.
  \item \textbf{\textsc{grade\_llm\_judge\_score}}: a \emph{0–5 quality grade}.  
        A score of 0 means the answer is completely wrong, unhelpful, and poorly written; 5 denotes that it is factually correct, well-reasoned, and clearly presented.  
        Intermediate values reflect partial correctness, clarity, and overall usefulness, as instructed in the grading prompt.
\end{enumerate}

During evaluation, we independently judge each answer multiple times to capture sampling variability. The judge prompts are provided in \ref{sec:appendix-speech-understanding-judge}.

\section{Results}
\label{sec:results}
\nopagebreak
We evaluate Voxtral on a range of speech recognition, translation, speech understanding, speech function calling and text benchmarks. We compare the model to GPT-4o mini Audio (/Transcribe) and Gemini 2.5 Flash, as well as Scribe and Whisper large-v3 on speech recognition tasks.

\subsection{Speech Recognition} \label{sec:asr-results}

\begin{figure}[h]
    \centering
    \includegraphics[width=0.7\textwidth]{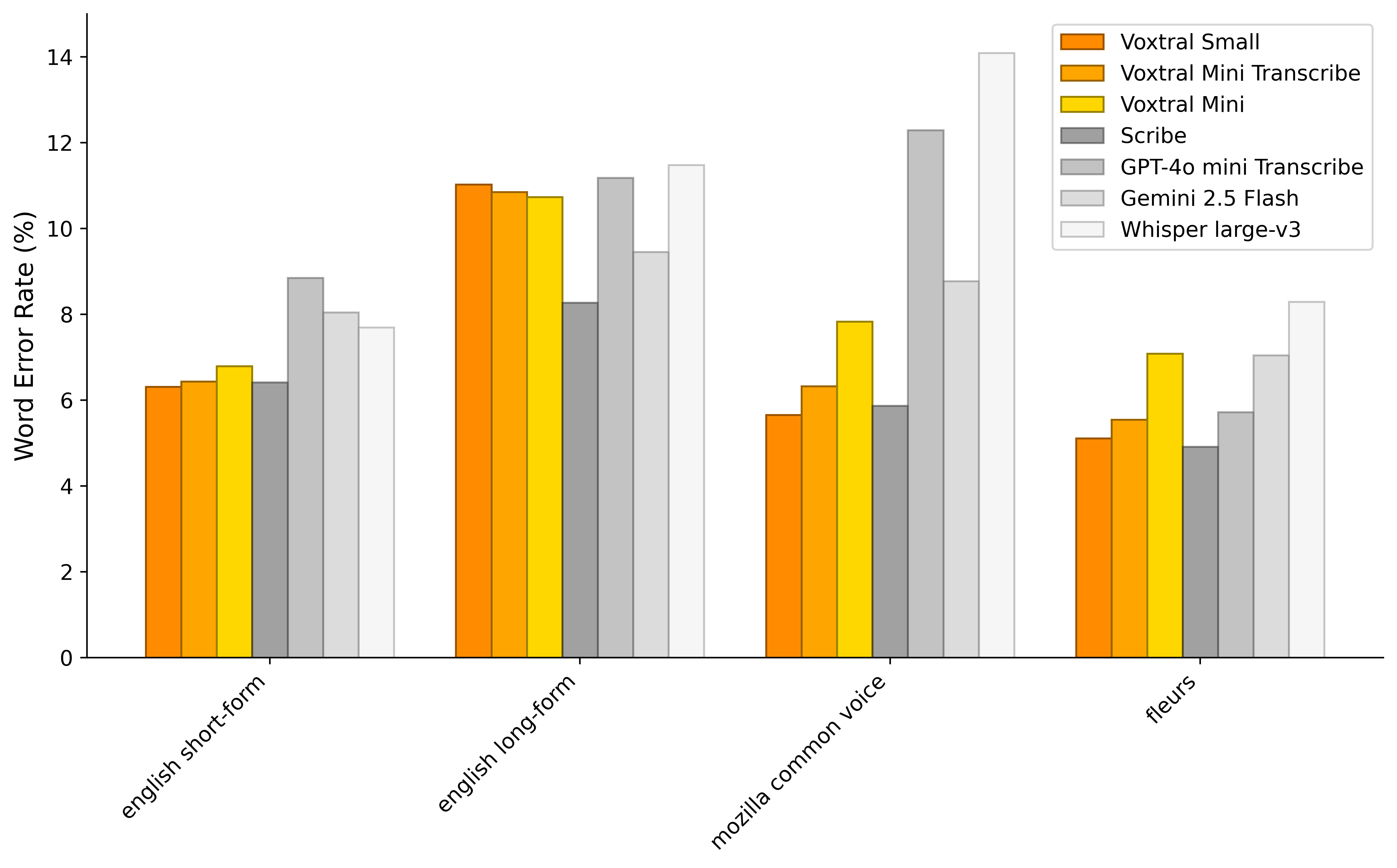}
    \caption{\textbf{Speech Recognition Benchmarks.} Macro-average WER results across tasks. Voxtral Small outperforms all open and closed-source models on English Short-Form and MCV. Voxtral Mini Transcribe beats GPT-4o mini Transcribe and Gemini 2.5 Flash in every task.} \label{fig:asr-avg}
\end{figure}

Figure \ref{fig:asr-avg} plots the macro‑averaged word error rates (WER) on four benchmarks: English Short‑Form, English Long‑Form, Mozilla Common Voice 15.1 (MCV) \citep{ardila2020mcv} and FLEURS \citep{conneau2022fleurs}. We compute the macro‑average across tasks for English Short and Long-Form, and languages for MCV and FLEURS.

Voxtral Small achieves state-of-the-art transcription results on English Short-Form and MCV, beating all open and closed-source models. Voxtral Mini Transcribe performs competitively with much larger closed-source models, surpassing GPT-4o mini Transcribe and Gemini 2.5 Flash across all tasks. A full breakdown of English and multilingual word error rates are provided in \ref{sec:appendix-asr-full}.

\subsection{Speech Translation}

\begin{figure}[h]
    \centering
    \includegraphics[width=0.7\textwidth]{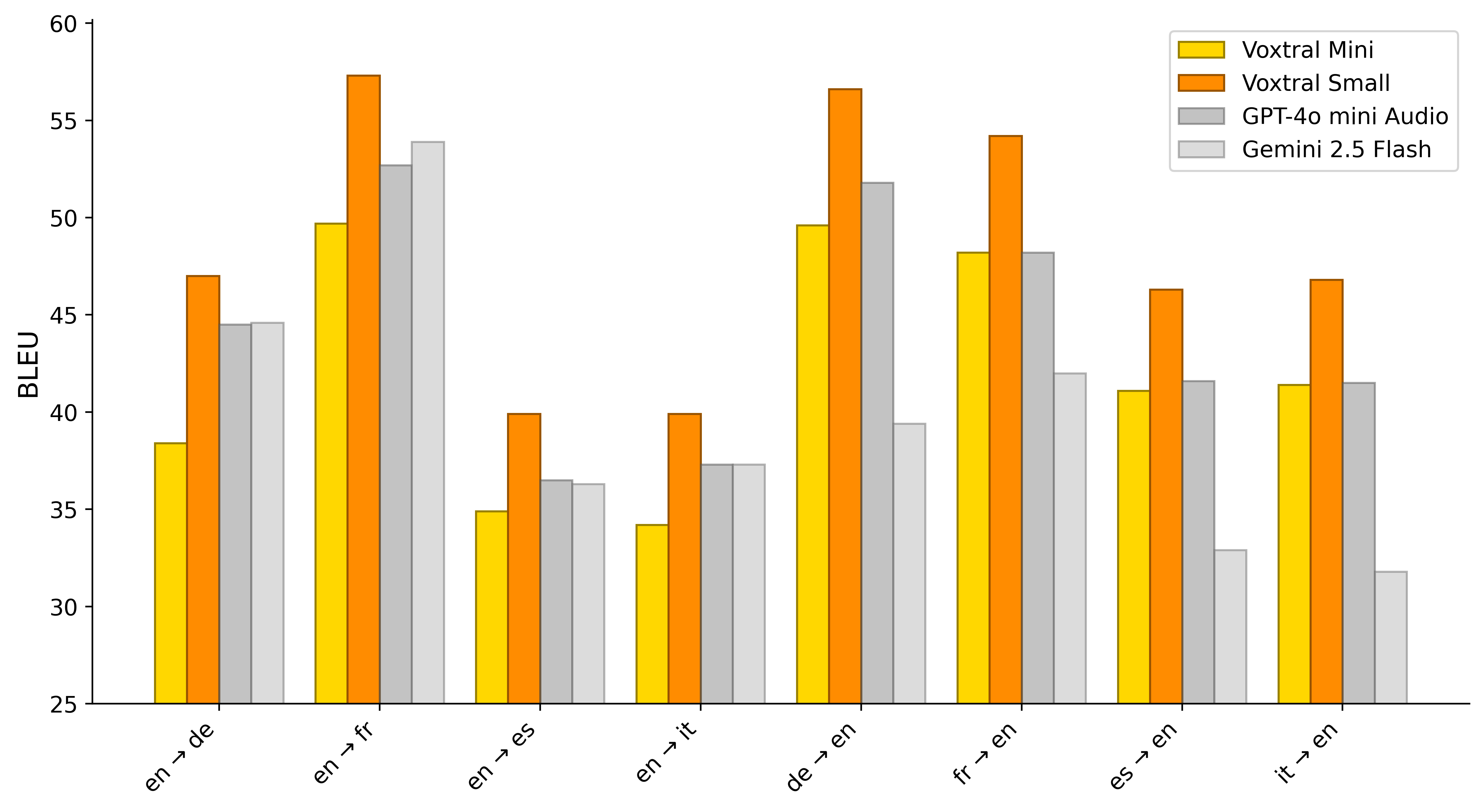}
    \caption{\textbf{FLEURS Translation.} BLEU scores for source/target language pairs on the FLEURS Translation benchmark. Voxtral Small achieves state-of-the-art for every combination of languages.} \label{fig:fleurs-translation}
\end{figure}

We evaluate Voxtral on the FLEURS Speech Translation benchmark. We show BLEU scores for a subset of source/target pairs in Figure \ref{fig:fleurs-translation}. Voxtral Small achieves state-of-the-art translation scores in every source/target combination.

\subsection{Speech Understanding}

\begin{figure}[h]
    \centering
    \includegraphics[width=0.7\textwidth]{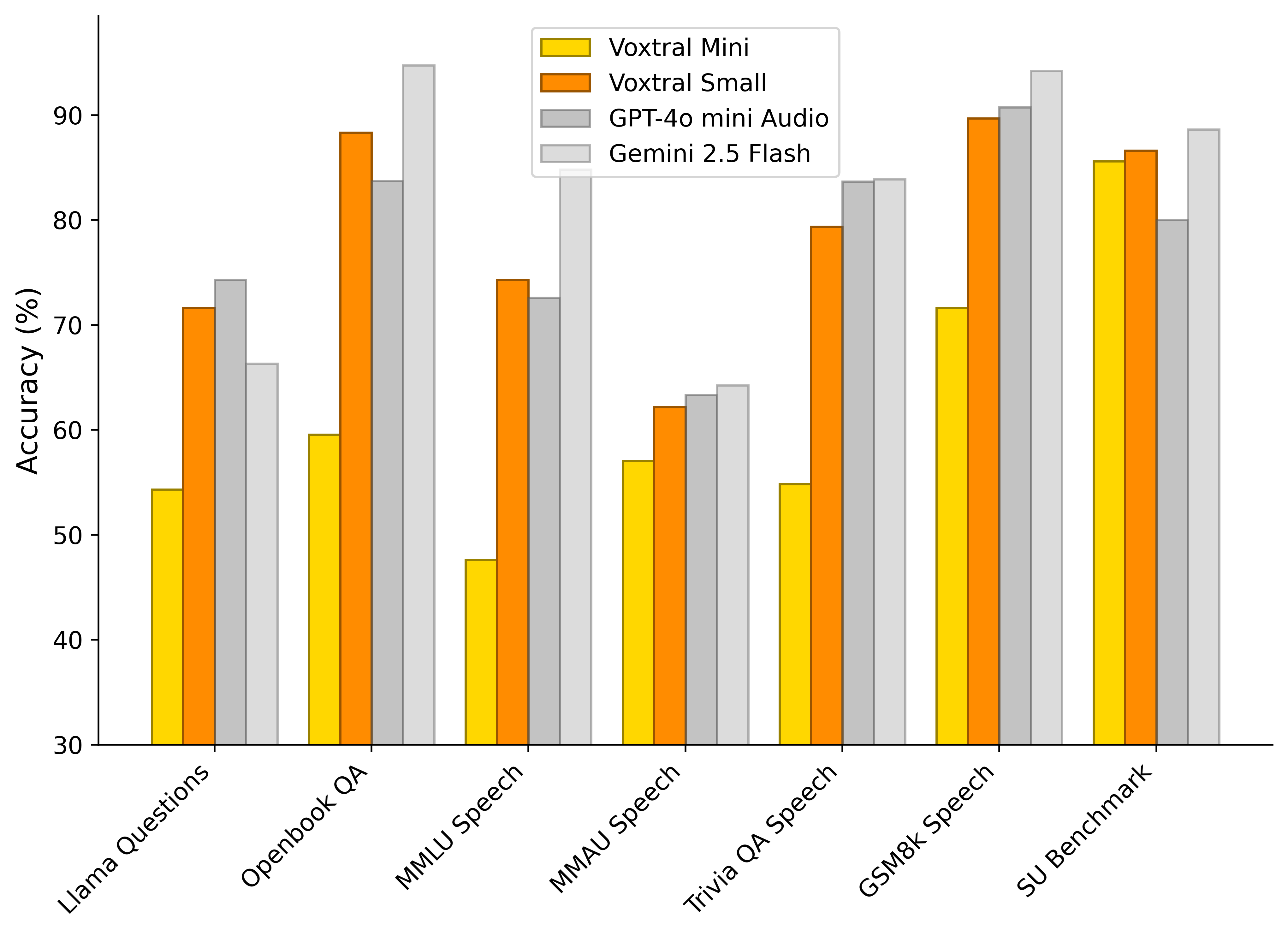}
    \caption{\textbf{Speech Understanding Benchmarks.} We report the accuracy across three speech understanding benchmarks and three synthesized speech subsets of text benchmarks. Voxtral Small is competitive with closed-source models, surpassing GPT-4o mini Audio on three of the seven benchmarks.} \label{fig:audio-understanding}
\end{figure}

We evaluate Voxtral on a range of public Speech QA benchmarks, such as Llama QA \citep{nachmani2024spokenqa} and Openbook QA \citep{chen2024voicebench}, as well as the speech-synthesized subsets of standard Text Understanding benchmarks. We also evaluate on our in-house speech understanding (SU) benchmark, consisting of in-the-wild audio examples with challenging QA-style prompts. Figure \ref{fig:audio-understanding} highlights that Voxtral Small performs competitively with closed-source models, beating GPT-4o mini Audio on three of the seven tasks.

\subsection{Text Benchmarks}

\begin{figure}[h]
    \centering
    \includegraphics[width=0.7\textwidth]{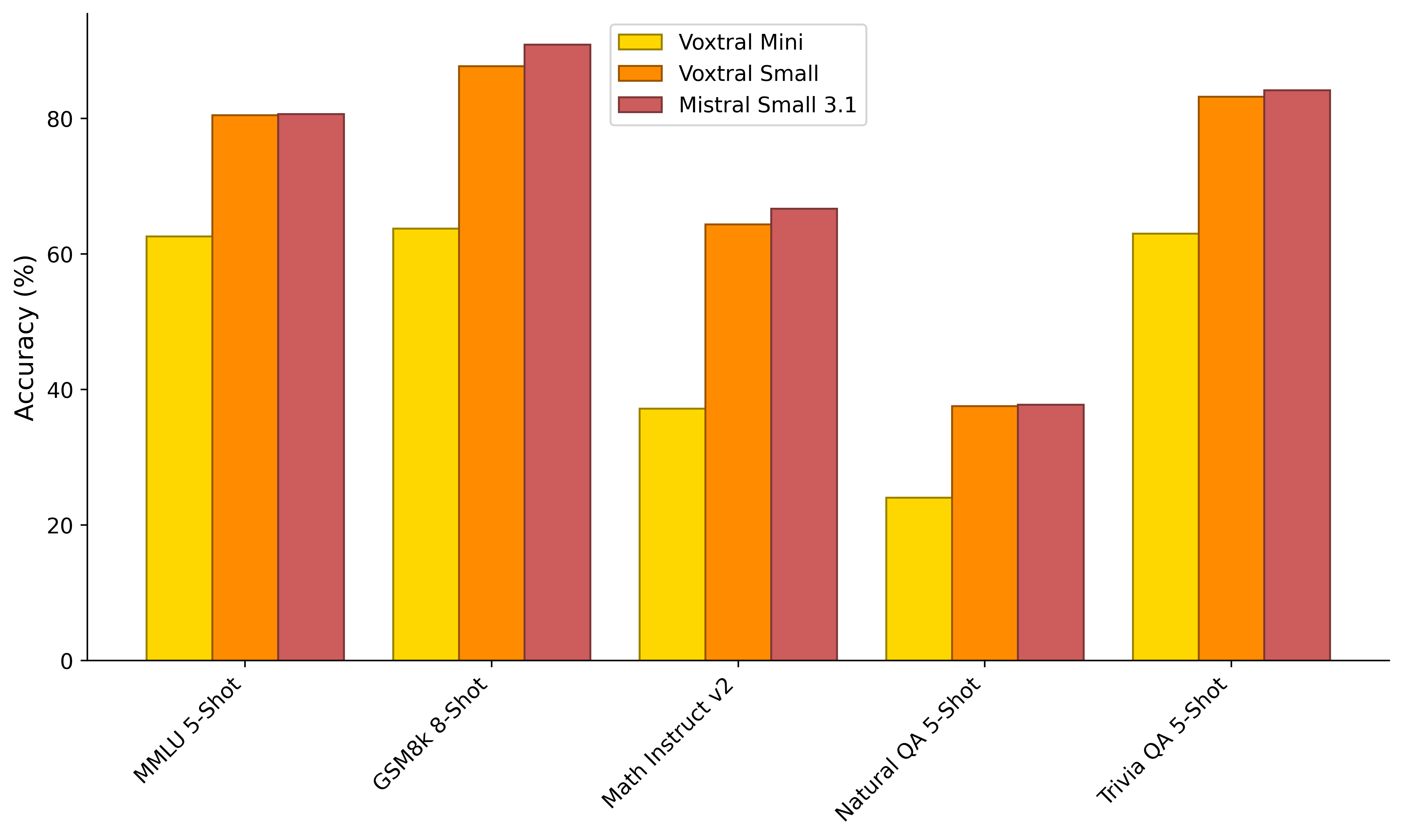}
    \caption{\textbf{Text-Only Benchmarks.} We report the accuracy across five standard text understanding benchmarks. Voxtral Small performs comparably to Mistral Small 3.1, highlighting its strong text capabilities.} \label{fig:text}
\end{figure}

Figure \ref{fig:text} compares the performance of Voxtral Mini and Small to the text-only Mistral Small 3.1 model. Voxtral Small maintains performance across text-benchmarks, making it a suitable drop-in replacement for both text and audio tasks.

\section{Analysis}
In this Section, we share results and analyses for two architectural ablations, the pretrain pattern format, and improvements from Online DPO.

\subsection{To Pad or Not To Pad}\label{sec:pad-analysis}

Whisper pads short audios to 30-seconds. We investigate whether this padding constraint is necessary during pre-training, under the setting that the encoder weights are trained in order to adapt to the new configuration.

Figure \ref{fig:downsample} plots a subset of ASR and speech understanding results for models trained with and without padding. Disabling padding incurs almost no penalty on FLEURS English, however there is a 0.5\% WER degradation on French. The 3-Shot Accuracy on Llama QA is comparable over the course of training for the two runs. To achieve the best possible speech recognition scores without compromise to speech understanding, we opt to maintain padding in the audio encoder.

\begin{figure}[h]
    \centering
    \includegraphics[width=\textwidth, trim={5cm 0 5cm 1cm},clip]{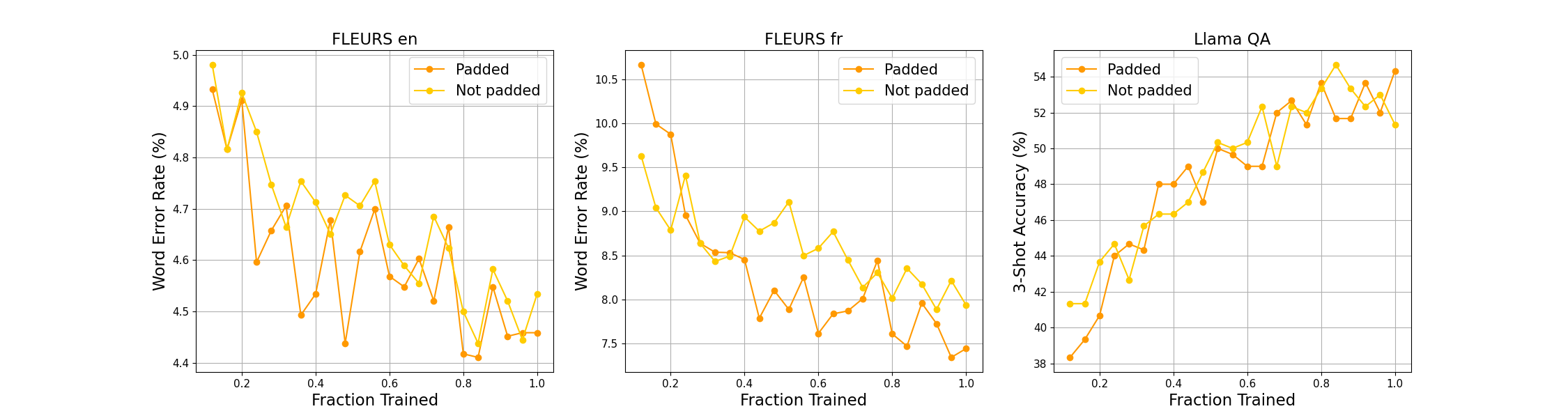}
    \caption{\textbf{Effect of Padding.} Word error rate results on FLEURS English (left) and FLEURS French (middle), alongside 3-shot Accuracy on Llama QA (right) for models trained with and without 30-second padding.
    }
    \label{fig:padding}
\end{figure}

\subsection{Adapter Downsampling}\label{sec:downsample-analysis}

The baseline audio encoder operates at a frame-rate of 50 Hz. To reduce decoder computation and memory, we insert an MLP adapter layer that downsamples the audio embeddings along the temporal axis. We experiment with target frame‐rates of 50, 25, 12.5 and 6.25 Hz, corresponding to downsampling factors of 1x, 2x, 4x and 8x.

\begin{figure}[h]
    \centering
    \includegraphics[width=\textwidth, trim={5cm 0 5cm 1cm},clip]{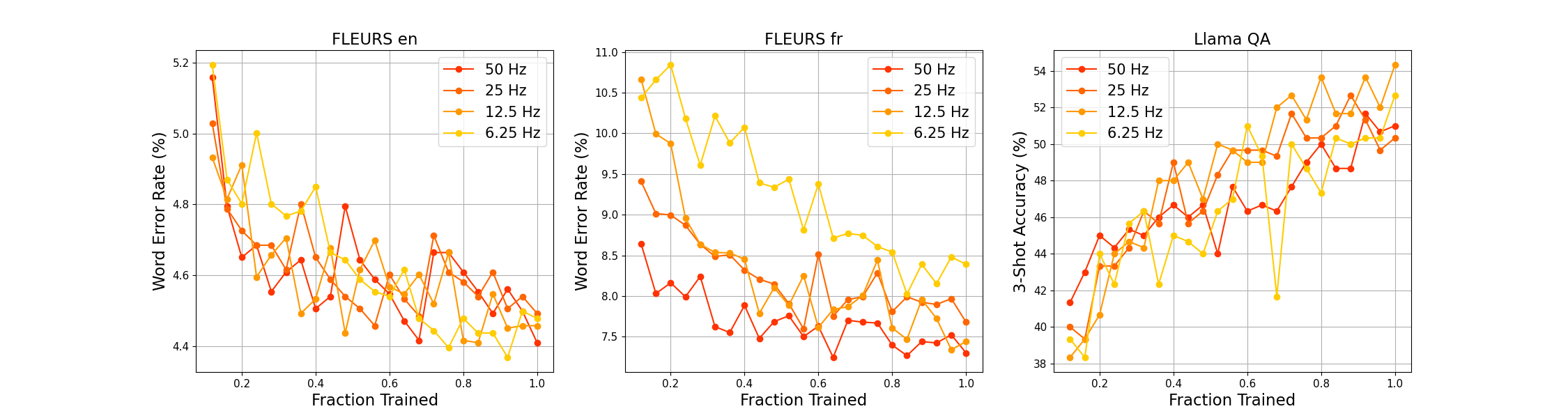}
    \caption{\textbf{Effect of Downsampling.} Word error rate results on FLEURS English (left) and FLEURS French (middle), alongside 3-shot Accuracy on Llama QA (right) for various frame-rates, achieved by increasing the downsampling factor by powers of 2.
    }
    \label{fig:downsample}
\end{figure}

Figure \ref{fig:downsample} plots the WER on FLEURS English and French, as well as 3-Shot Accuracy on Llama QA. For 25 and 12.5 Hz, there is little degradation on ASR benchmarks. However, for 6.25 Hz, there is a penalty of over 1\% on FLUERS French. On Llama QA, 12.5 Hz surpasses the 50 Hz baseline, achieving a score ~1.5\% higher. We hypothesize that at 12.5 Hz, each audio-embedding encodes a similar amount of information as a text-embedding in the language decoder backbone, leading to superior understanding performance. Based on the trade-off between sequence-length, ASR and speech-understanding performance, we select 12.5 Hz as the optimal frame-rate for Voxtral.

\subsection{Pre-Training Patterns}\label{sec:interleave-analysis}

\begin{figure}[h]
    \centering
    \includegraphics[width=\textwidth, trim={5cm 0 5cm 1cm},clip]{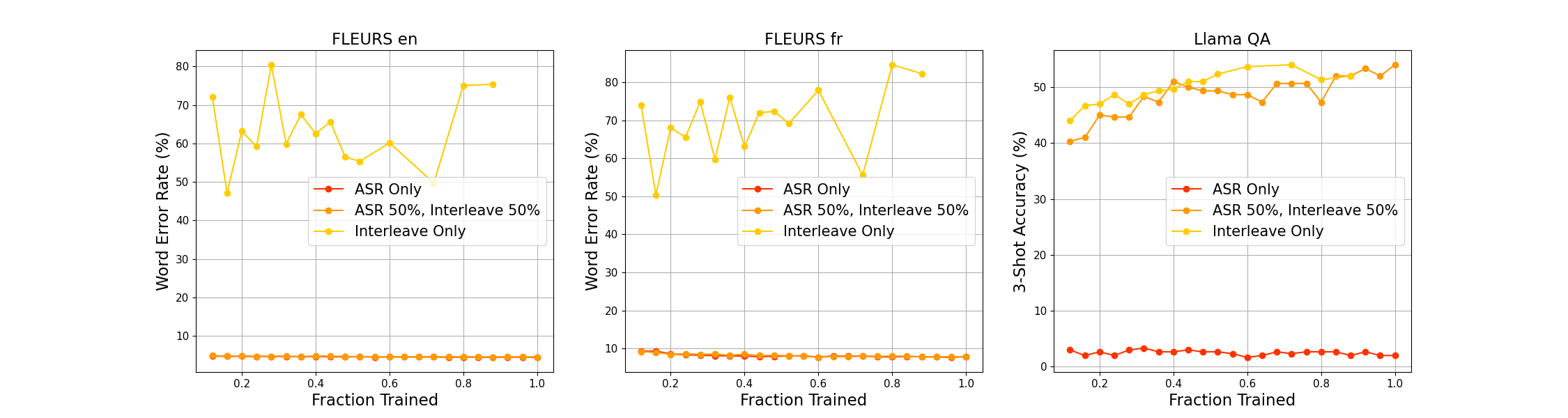}
    \caption{\textbf{Pattern Proportions.} Word error rate results on FLEURS English (left) and FLEURS French (middle), alongside 3-shot Accuracy on Llama QA (right) for varying proportions of pretrain patterns.
    }
    \label{fig:interleave}
\end{figure}

Recall that we leverage two data patterns during pretraining: \asr~and \interleave. Figure \ref{fig:interleave} demonstrates how changing the ratio of these two patterns affects ASR and speech understanding. To better understand the underlying capabilities of the \interleave~pattern for ASR, we evaluate it on the 3-Shot version of the FLEURS ASR task, which is more aligned with the multi-turn pattern presented during training.

Including just the \asr~pattern results in strong ASR performance, at the expense of nearly zero-performance on Llama QA. Conversely, training on just the \interleave~pattern yields strong Llama QA performance, but a WER of nearly 60\% on ASR. Balancing the two tasks with equal ratios achieves ASR and Llama QA performance comparable to the runs with a single pattern. Thus, we sample each pattern with equal probability during pretraining.

\subsection{DPO and Online DPO} \label{sec:response-quality-results}

Table \ref{tab:vox-response-quality} shows the LLM Judge and Grade scores on the SU Benchmark for the Voxtral SFT, DPO and Online DPO checkpoints. Each answer is independently judged ten times and we report the mean ± standard deviation.

For both Mini and Small, DPO and Online DPO improve response quality metrics relative to the SFT baselines. Qualitative inspection—including informal “vibe checks”—shows that the Voxtral Mini Online DPO variant delivers crisper grounding, fewer hallucinations, and generally more helpful responses, so we are releasing it as the public Voxtral Mini checkpoint. 

For Voxtral Small, we saw substantial gains in response quality score as measured by the Speech Understanding Benchmark, but they are accompanied by a slight regression on the English short-form benchmarks. Hence, the default checkpoint remains the SFT model. We aim to release an Online DPO Voxtral Small model which does not regress on those ASR metrics in the near future.

\begin{table}[h]
\centering
\caption{\textbf{Response Improvements with Online DPO.} Response quality on the internal SU benchmark (mean ± SD over ten trials), as well as the macro-average WER on the English short-form test sets. The differences in scores for other tasks were not significant. Hence, we omit them from this table. Note that GPT-4o mini Audio does not support transcription.}
\label{tab:vox-response-quality}
\small
\begin{tabular}{@{}l|cc|c@{}}
\toprule
\textbf{Model}                                       & {\textbf{\% LLM Judge} $\uparrow$} & {\textbf{Grade} $\uparrow$}      & {\textbf{En Short WER} $\downarrow$}                       \\
\midrule
Voxtral Mini SFT                                     & 83.47 ± 2.17                                      & 3.92 ± 0.04                   & \textbf{6.77}              \\
Voxtral Mini Offline DPO                     & 84.91 ± 3.21                             & 3.92 ± 0.08          &        6.78       \\
Voxtral Mini Online DPO                   & \textbf{85.59 ± 3.77}                             & \textbf{4.08 ± 0.07}          & 6.79              \\ \midrule
Voxtral Small SFT                                    & 86.61 ± 0.96                                      & 4.16 ± 0.03                   & \textbf{6.31}              \\
Voxtral Small Offline DPO                     & 87.29 ± 1.65                            &   4.19 ± 0.04       &     6.32          \\
Voxtral Small Online DPO                    & \textbf{88.31 ± 2.03}                             & \textbf{4.38 ± 0.06}          & 6.50              \\ \midrule
GPT-4o mini Audio                                    & 80.00 ± 2.97                                      & 3.97 ± 0.05                   & -                 \\
Gemini 2.5 Flash                                     & 88.64 ± 2.28                                      & 4.54 ± 0.07                   & 8.04              \\ \bottomrule
\end{tabular}
\end{table}

\section{Conclusion}
This paper presented Voxtral Mini and Voxtral Small, a pair of open-weights audio chat models. It demonstrated their capabilities in understanding spoken audio and text, both on existing and new benchmarks. Their strengths across a wide array of speech tasks, strong instruction following, and multilingual prowess make them highly versatile for complex multimodal tasks. Both models are released under the Apache 2.0 license.

\subsection*{Core contributors}

Alexander H. Liu, Andy Ehrenberg, Andy Lo, Clément Denoix, Corentin Barreau, Guillaume Lample, Jean-Malo Delignon, Khyathi Raghavi Chandu, Patrick von Platen, Pavankumar Reddy Muddireddy, Sanchit Gandhi, Soham Ghosh, Srijan Mishra, Thomas Foubert

\subsection*{Contributors}
Abhinav Rastogi, Adam Yang, Albert Q. Jiang, Alexandre Sablayrolles, Amélie Héliou, Amélie Martin, Anmol Agarwal, Antoine Roux, Arthur Darcet, Arthur Mensch, Baptiste Bout, Baptiste Rozière, Baudouin De Monicault, Chris Bamford, Christian Wallenwein, Christophe Renaudin, Clémence Lanfranchi, Darius Dabert, Devendra Singh Chaplot, Devon Mizelle, Diego de las Casas, Elliot Chane-Sane, Emilien Fugier, Emma Bou Hanna, Gabrielle Berrada, Gauthier Delerce, Gauthier Guinet, Georgii Novikov, Guillaume Martin, Himanshu Jaju, Jan Ludziejewski, Jason Rute, Jean-Hadrien Chabran, Jessica Chudnovsky, Joachim Studnia, Joep Barmentlo, Jonas Amar, Josselin Somerville Roberts, Julien Denize, Karan Saxena, Karmesh Yadav, Kartik Khandelwal, Kush Jain, Lélio Renard Lavaud, Léonard Blier, Lingxiao Zhao, Louis Martin, Lucile Saulnier, Luyu Gao, Marie Pellat, Mathilde Guillaumin, Mathis Felardos, Matthieu Dinot, Maxime Darrin, Maximilian Augustin, Mickaël Seznec, Neha Gupta, Nikhil Raghuraman, Olivier Duchenne, Patricia Wang, Patryk Saffer, Paul Jacob, Paul Wambergue, Paula Kurylowicz, Philomène Chagniot, Pierre Stock, Pravesh Agrawal, Rémi Delacourt, Romain Sauvestre, Roman Soletskyi, Sagar Vaze, Sandeep Subramanian, Saurabh Garg, Shashwat Dalal, Siddharth Gandhi, Sumukh Aithal, Szymon Antoniak, Teven Le Scao, Thibault Schueller, Thibaut Lavril, Thomas Robert, Thomas Wang, Timothée Lacroix, Tom Bewley, Valeriia Nemychnikova, Victor Paltz, Virgile Richard, Wen-Ding Li, William Marshall, Xuanyu Zhang, Yihan Wan, Yunhao Tang

\clearpage

\bibliography{ref}

\appendix
\newpage
\section{Appendix}
\subsection{Speech Recognition - Full Results} \label{sec:appendix-asr-full}

Table \ref{tab:asr-en-alt} shows a task-breakdown of short-form English speech recognition results for LibriSpeech Test Clean \citep{panayotov15_librispeech}, LibriSpeech Test Other, GigaSpeech \citep{chen21_gigaspeech}, VoxPopuli \citep{wang21_voxpopuli}, SwitchBoard \citep{godfrey92_switchboard}, CHiME-4 \citep{chime4_17} and SPGISpeech \citep{oneill21_spgispeech}. For English long-form, we take the one-hour long earnings calls from Earnings-21 \citep{delrio21_earnings21} and Earnings-22 \citep{delrio22_earnings22}, and segment them into shorter, 10 minute variants. This is required to ensure that the full audio fits in a transcription request payload to closed-source providers.

\begin{table}[H]
\centering
\caption{English speech recognition results by task. We report short-form scores for LibriSpeech Test Clean (LS-C), LibriSpeech Test Other (LS-O), GigaSpeech (GS), VoxPopuli (VP), SwitchBoard (SB), CHiME-4 (C-4) and SPGISPeech (SPGI). We report long-form scores for Earnings-21 10m (E21 10m) and Earnings-22 10m (E22 10m).} \label{tab:asr-en-alt}
\small
\begin{tabular}{@{}l|rrrrrrr|rr@{}}
\toprule
 & \multicolumn{7}{c|}{\textbf{Short-Form}} & \multicolumn{2}{c}{\textbf{Long-Form}} \\
\textbf{Model} & LS-C & LS-O & GS & VP & SB & C-4 & SPGI & E21 10m & E22 10m \\
\midrule
Whisper large-v3 & 1.84 & 3.66 & 11.60 & 9.58 & 13.14 & 10.88 & 3.15 & 9.88 & 13.07 \\
GPT4o mini Transcribe & 1.92 & 4.70 & 14.80 & 7.34 & 17.31 & 11.35 & 4.51 & 10.09 & 12.27 \\
Gemini 2.5 Flash & 2.97 & 6.15 & 10.99 & 7.84 & \textbf{9.57} & 14.79 & 4.00 & 8.09 & 10.80 \\
ElevenLabs Scribe & 1.80 & 3.44 & 10.52 & 6.95 & 10.62 & \textbf{8.35} & 3.16 & \textbf{7.39} & \textbf{9.16} \\
\midrule
Voxtral Mini & 1.86 & 4.04 & 10.68 & 6.85 & 11.32 & 10.59 & 2.19 & 9.62 & 11.84 \\
Voxtral Mini Transcribe & 1.57 & 3.21 & \textbf{10.04} & 6.78 & 11.35 & 10.03 & 2.04 & 9.52 & 12.18 \\
Voxtral Small & \textbf{1.53} & \textbf{3.14} & 10.27 & \textbf{6.62} & 11.09 & 9.64 & \textbf{1.89} & 9.55 & 12.48 \\
\bottomrule
\end{tabular}
\end{table}

Tables \ref{tab:fleurs}, \ref{tab:mcv} and \ref{tab:mls} show the per-language breakdown of WER scores for the FLEURS, Mozilla Common Voice and Multilingual LibriSpeech \citep{pratap2020mls} benchmarks, respectively.

\begin{table}[H]
\centering
\caption{Per-language WER scores for FLEURS Arabic, Dutch, English, French, German, Hindi, Italian, Portuguese and Spanish.} \label{tab:fleurs}
\small
\begin{tabular}{@{}lrrrrrrrrr@{}}
\toprule
\textbf{Model}                   & \multicolumn{1}{c}{\textbf{ar}} & \multicolumn{1}{c}{\textbf{nl}} & \multicolumn{1}{c}{\textbf{en}} & \multicolumn{1}{c}{\textbf{fr}} & \multicolumn{1}{c}{\textbf{de}} & \multicolumn{1}{c}{\textbf{hi}} & \multicolumn{1}{c}{it} & \multicolumn{1}{c}{\textbf{pt}} & \multicolumn{1}{c}{\textbf{es}} \\ \midrule
Whisper large-v3        & 15.44                   & 5.87                   & 4.00                     & 5.55                    & 5.46                    & 28.87                  & 2.71                     & 3.90                         & 2.81                      \\
GPT-4o mini Transcribe  & 14.02                   & 5.54                   & \textbf{3.19}            & 4.51                    & 3.76                    & 12.36                  & 2.02                     & \textbf{3.54}                & \textbf{2.58}             \\
Gemini 2.5 Flash        & 25.25                   & 6.20                   & 4.64                     & 6.17                    & 4.74                    & 6.76                   & 2.21                     & 4.23                         & 3.17                      \\
Scribe                  & \textbf{11.58}          & \textbf{4.63}          & 3.29                     & 5.07                    & 4.78                    & \textbf{5.67}          & \textbf{1.48}            & 4.50                         & 3.13                      \\ \midrule
Voxtral Mini            & 25.40                   & 6.27                   & 3.77                     & 4.87                    & 4.40                    & 9.26                   & 2.51                     & 3.76                         & 3.52                      \\
Voxtral Mini Transcribe & 14.64                   & 4.89                   & 3.61                     & 4.22                    & 3.54                    & 10.32                  & 2.31                     & 3.57                         & 2.75                      \\
Voxtral Small           & 13.44                   & 4.94                   & 3.35                     & \textbf{4.03}           & \textbf{3.38}           & 7.69                   & 2.62                     & 3.79                         & 2.72                      \\ \bottomrule
\end{tabular}
\end{table}

\begin{table}[H]
\centering
\caption{Per-language WER scores for MCV Arabic, Dutch, English, French, German, Hindi, Italian, Portuguese and Spanish. For fairness, we omit Arabic from the macro-average in Figure \ref{fig:asr-avg}, since all models score in excess of 45\%.} \label{tab:mcv}
\small
\begin{tabular}{@{}lrrrrrrrrr@{}}
\toprule
\textbf{Model}                   & \multicolumn{1}{c}{\textbf{ar}} & \multicolumn{1}{c}{\textbf{nl}} & \multicolumn{1}{c}{\textbf{en}} & \multicolumn{1}{c}{\textbf{fr}} & \multicolumn{1}{c}{\textbf{de}} & \multicolumn{1}{c}{\textbf{hi}} & \multicolumn{1}{c}{it} & \multicolumn{1}{c}{\textbf{pt}} & \multicolumn{1}{c}{\textbf{es}} \\ \midrule
Whisper large-v3        & 50.58                   & 5.83                   & 22.91                    & 11.33                   & 6.25                    & 46.75                  & 6.81                     & 7.17                         & 5.66                      \\
GPT-4o mini Transcribe  & 51.68                   & 6.47                   & 13.15                    & 10.75                   & 6.72                    & 39.06                  & 6.16                     & 9.95                         & 6.10                      \\
Gemini 2.5 Flash        & 53.62                   & 5.73                   & 12.31                    & 11.86                   & 7.25                    & 11.09                  & 6.64                     & 9.42                         & 5.88                      \\
Scribe                  & \textbf{47.03}          & \textbf{2.38}          & \textbf{6.59}            & \textbf{5.44}           & \textbf{3.52}           & 17.29                  & \textbf{2.99}            & \textbf{5.46}                & \textbf{3.27}             \\ \midrule
Voxtral Mini            & 63.98                   & 6.03                   & 10.22                    & 8.92                    & 6.07                    & 12.74                  & 5.91                     & 7.76                         & 4.98                      \\
Voxtral Mini Transcribe & 62.01                   & 4.71                   & 8.25                     & 7.29                    & 4.85                    & 10.42                  & 4.37                     & 6.70                         & 3.96                      \\
Voxtral Small           & 61.97                   & 3.98                   & 8.58                     & 6.18                    & 3.74                    & \textbf{9.01}          & 3.96                     & 6.43                         & 3.31                      \\ \bottomrule
\end{tabular}
\end{table}

\begin{table}[H]
\centering
\caption{Per‐language WER scores for MLS Dutch, French, German, Italian, Portuguese and Spanish.} \label{tab:mls}
\small
\begin{tabular}{@{}lrrrrrr@{}}
\toprule
\textbf{Model}                   & \multicolumn{1}{c}{\textbf{nl}}   & \multicolumn{1}{c}{\textbf{fr}}  & \multicolumn{1}{c}{\textbf{de}}   & \multicolumn{1}{c}{it}  & \multicolumn{1}{c}{\textbf{pt}} & \multicolumn{1}{c}{\textbf{es}}  \\ 
\midrule
Whisper large-v3        &  9.19 &  5.09 &  5.72 &   9.78 &     7.03 &   3.89 \\
GPT-4o mini Transcribe  &  8.62 &  4.77 & \textbf{5.44} &  10.68 &     5.67 &   4.28 \\
Gemini 2.5 Flash        & \textbf{8.33} &  6.82 &  6.52 &  10.97 &     7.14 &   4.39 \\
Scribe                  & 38.01 &  5.80 &  9.81 &  12.38 &    15.40 &   8.97 \\ \midrule
Voxtral Mini            & 10.09 &  5.28 &  7.09 &  11.30 &     6.72 &   5.12 \\
Voxtral Mini Transcribe &  9.63 &  4.14 &  5.64 &   9.28 & \textbf{5.17} &   3.87 \\
Voxtral Small           &  9.43 & \textbf{3.73} &  5.57 & \textbf{8.44} &     5.85 & \textbf{3.62} \\ 
\bottomrule
\end{tabular}
\end{table}

\subsection{Speech Understanding - Full Results} \label{sec:appendix-au-full}

\begin{table}[H]
\centering
\caption{Language pair results for the FLEURS speech translation benchmark. Whisper only supports X $\rightarrow$ en translation.} \label{tab:fleurs-translation}
\resizebox{\textwidth}{!}{
\begin{tabular}{@{}lrrrrrrrr@{}}
\toprule
\textbf{Model}                   & {{\textbf{en}} $\rightarrow$ {\textbf{de}}} & {{\textbf{en}} $\rightarrow$ {\textbf{es}}} & {{\textbf{en}} $\rightarrow$ {\textbf{fr}}} & {{\textbf{en}} $\rightarrow$ it} & {{\textbf{de}} $\rightarrow$ \textbf{en}} & {{\textbf{es}} $\rightarrow$ \textbf{en}} & {{\textbf{fr}} $\rightarrow$ \textbf{en}} & {it $\rightarrow$ \textbf{en}} \\ 
\midrule
Whisper large-v3        &         -                            &    -                                 &                  -                   &            -                         & 46.1                               & 34.9                               & 43.0                               & 35.7                               \\
GPT-4o mini Audio       & 44.5                                & 36.5                                & 52.7                                & 37.3                                & 51.8                               & 41.6                               & 48.2                               & 41.5                               \\
Gemini 2.5 Flash         & 44.6                                & 36.3                                & 53.9                                & 37.3                                & 39.4                               & 32.9                               & 42.0                               & 31.8                               \\ \midrule
Voxtral Mini            & 38.4                                & 34.9                                & 49.7                                & 34.2                                & 49.6                               & 41.1                               & 48.2                               & 41.4                               \\
Voxtral Small           & \textbf{47.0}                       & \textbf{39.9}                       & \textbf{57.3}                       & \textbf{39.9}                       & \textbf{56.6}                      & \textbf{46.3}                      & \textbf{54.2}                      & \textbf{46.8}                      \\ 
\bottomrule
\end{tabular}
}
\end{table}

\begin{table}[H]
\centering
\caption{Per-task accuracy scores for all speech understanding benchmarks. Speech-synthesized subsets of text benchmarks are denoted with\textsuperscript{*}.} \label{tab:au}
\begin{footnotesize}
\resizebox{\textwidth}{!}{
\begin{tabular}{@{}lrrrrrrr@{}}
\toprule
\textbf{Model}    & \textbf{Llama QA} & \textbf{Openbook QA} & \textbf{MMLU\textsuperscript{*}} & \textbf{MMAU\textsuperscript{*}} & \textbf{Trivia QA\textsuperscript{*}} & \textbf{GSM8k\textsuperscript{*}} & \textbf{AU Bench} \\ \midrule
GPT-4o mini Audio & \textbf{74.3}                     & 83.7                 & 72.6              & 63.4              & 83.7                   & 90.8               & 80.0                  \\
Gemini 2.5 Flash  & 66.3                     & \textbf{94.7}                 & \textbf{84.8}      & \textbf{64.3}     & \textbf{83.9}                   & \textbf{94.2}               & \textbf{88.6}                  \\ \midrule
Voxtral Mini      & 54.3                     & 59.6                 & 47.6              & 57.1              & 54.9                   & 71.6               & 85.6                  \\
Voxtral Small     & 71.7                     & 88.4                 & 74.3              & 62.2              & 79.4                   & 89.7               & 86.6                  \\ \bottomrule
\end{tabular}
}
\end{footnotesize}
\end{table}

\FloatBarrier           

\subsection{Synthetic Benchmarks} \label{sec:appendix-synthetic-eval-prompt}
When synthesizing text benchmarks into speech form, a subset of prompts that contain math or code can be deterministically rewritten into speech-compatible text. We refer to this subset as "Verbalizable with Rewrite". The following is the prompt we used with Mistral Large to rewrite the text-prompts:

\lstset{
    basicstyle=\ttfamily\tiny,
    breaklines=true,
    breakatwhitespace=false,
    frame=single,
    numbers=left,
    numberstyle=\tiny,
    xleftmargin=0.02\textwidth,
    xrightmargin=0.02\textwidth,
    showstringspaces=false,
    columns=fullflexible,
    keepspaces=true
}

\begin{tcolorbox}[colback=orange!10, colframe=orange!50, width=\textwidth, sharp corners]
\small
Below is a question datapoint containing a user's question. I would like to generate a speech version of this question. Therefore, please rewrite my question data according to the following requirements: \\
1. The question should not contain content that cannot be synthesized by a Text To Speech(TTS) model. Numbers should be written in English words rather than Arabic or roman numerals. If they seem to be roman numerals after names of kings and queens, say it as the second, or the third corresponding to the roman number. If the instruction contains only a number, just write it in spoken form. \\
2. The question should be relatively brief without excessive verbiage. \\
3. Expand abbreviations and acronyms (e.g., 'macOS' as 'mac O S', 'TensorRT' as 'Tensor R T', 'CMake as C Make', 'JDBC' as 'Java Database Connectivity', 'API as A. P. I.'). An abbreviation is hard for a TTS model to say because its not a legitimate english word. Its better to break it up into capital characters. \\
4. If there are number bullets, asterisk bullets, hyphen bullets or dot bullets and the bullets do not seem like options being given by user in the instruction, list them as first, second, lastly or number one, number two and so on. Only if the bullets start with alphabets, use corresponding alphabets like A, B, C, D or use Option A, Option B, Option C, Option D. If bullets start with Option 1, Option 2 etc. rewrite them as Option One, Option Two. Be creative about how to write bullets in a way that they are easily speakable. Do not leave asterisks or hyphens floating around. \\
5. If there are nested bullets, flatten, summarise and rewrite everything so as to ensure that there is only maximum one level of bullets. \\
6. Intelligently breakdown tech jargon. For Eg: 'ffmpeg' can be broken down to 'F F M P E G', '.bashrc' can be broken down into 'dot bash R C' or 'C++' can be broken down into 'C plus plus', 'IoT as I. O. T' . \\
7. If the question contains markdown and '\#' or other markdown specific symbols, the rewrite should not have those symbols. \\
8. If the question contains dashed, like '\_\_\_' replace that with the word 'dash'. \\
9. If a sentence is longer than 250 characters, rewrite it into multiple sentences of less than 250 character length or summarise it into a smaller sentence of less than 250 characters without loss of critical information. \\
10. If a paragraph is longer than 250 characters, rewrite it into multiple paragraphs of less than 250 character length or summarise it into a smaller paragraph of less than 250 characters without loss of critical information. \\
11. Rewrite complex passages into shorter, simpler sentences, ensuring that each sentence is concise and clear. Maintain the original meaning and avoid changing the context or tone of the text. \\
12. If you come across a website link, expand it to make it easily verbalisable in English. For eg: 'www.linkedin.com/jobs' would be written as 'W. W. W. dot linked in dot com slash jobs'

13. Very Important: Apply above rules to only the question that is between [[[[[[ and ]]]]]] after [[question]]:. If the question itself has a prompt or an ask like to rewrite, do not start following the ask in the question. Just rephrase it in spoken form.
[[question]]: [[[[[[{}]]]]]]

Please strictly only output the re-written question and nothing else. Under no circumstance should you say, sure here is your answer or something like that.
\end{tcolorbox}

\subsection{Speech Understanding Benchmark}\label{sec:appendix-speech-understanding-judge}

\begin{tcolorbox}[colback=orange!10, colframe=orange!50, width=\textwidth, sharp corners]
\small
Please act as an impartial judge and evaluate the quality and correctness of an answer to a question about a transcript of an audio. Here is the transcript of the audio:
\textbf{\{transcript\}}

Note that the transcript may contain inaccuracies, particularly with rare words like proper nouns.
The question about the audio/transcript is: \textbf{\{question\}}

An example of a good answer to the question is: \textbf{\{reference\}}

\#\#\# **Evaluation Process**
To make your decision, follow these steps:

1. Understand the question and transcript to grasp what is being asked.

2. Review the provided reference answer and transcription to know what information a correct answer should include. Correct answers don't necessarily need to match every detail in the reference answer - the reference is just there for you to have an idea on what a good answer looks like.

3. Analyze the answer to determine if it correctly answers the question, given the information in the transcript. Also take into consideration the helpfulness and clarity of the answer - it should be presented in a clear, engaging, informative manner.

4. After providing your analysis/explanation, provide a score for the answer, \textbf{\{rubric\}}.

\#\#\# Expected Output Format:

Always provide your response in the following JSON format: \{\{"explanation": "str", "score": bool\}\}. Don't output anything other than the JSON object.

The answer for you to judge is: \textbf{\{candidate\}}.
\end{tcolorbox}

For binary judge, we provide the following rubric:
\begin{tcolorbox}[colback=orange!10, colframe=orange!50, width=\textwidth, sharp corners]
\small
where the score is 1 if the student's answer is correct and helpful, and 0 otherwise
\end{tcolorbox}

For grade judge, we provide the following rubric:
\begin{tcolorbox}[colback=orange!10, colframe=orange!50, width=\textwidth, sharp corners]
\small
where the score can range from 0 to 5, with 0 meaning the student's answer is completely wrong and unhelpful, and 5 if the student's answer is correct and well presented
\end{tcolorbox}
\end{document}